\documentclass[useAMS,usenatbib,usegraphicx,onecolumn,fleqn]{mn2e}
\usepackage{amsmath}
\usepackage{times}

\title{The Skewness of the Aperture Mass Statistic}

\author[M. Jarvis \etal]
{M. Jarvis,$^1$ G. Bernstein,$^1$ B. Jain$^1$ \\
$^1$ Department of Physics and Astronomy, University of Pennsylvania,
Philadelphia, PA 19104 }

\def\eqq#1{Equation~(\ref{#1})}
\newcommand\etal{{\rm et al.}}
\newcommand{\map}{\mbox{$M_{\rm ap}$}}
\newcommand{\mx}{\mbox{$M_{\times}$}}
\newcommand{\mapsq}{\mbox{$\langle M_{\rm ap}^2\rangle$}}
\newcommand{\mapmx}{\mbox{$\langle M_{\rm ap} M_{\times}\rangle$}}
\newcommand{\mxsq}{\mbox{$\langle M_{\times}^2\rangle$}}
\newcommand{\msq}{\mbox{$\langle M^2\rangle$}}
\newcommand{\mmc}{\mbox{$\langle M M^*\rangle$}}
\newcommand{\mapcb}{\mbox{$\langle M_{\rm ap}^3\rangle$}}
\newcommand{\mapsqmx}{\mbox{$\langle M_{\rm ap}^2 M_{\times}\rangle$}}
\newcommand{\mapmxsq}{\mbox{$\langle M_{\rm ap} M_{\times}^2\rangle$}}
\newcommand{\mxcb}{\mbox{$\langle M_{\times}^3\rangle$}}
\newcommand{\mcb}{\mbox{$\langle M^3\rangle$}}
\newcommand{\msqmc}{\mbox{$\langle M^2 M^*\rangle$}}
\newcommand{\mapqd}{\mbox{$\langle M_{\rm ap}^4\rangle$}}
\newcommand{\mapcbmx}{\mbox{$\langle M_{\rm ap}^3 M_{\times}\rangle$}}
\newcommand{\mapsqmxsq}{\mbox{$\langle M_{\rm ap}^2 M_{\times}^2\rangle$}}
\newcommand{\mapmxcb}{\mbox{$\langle M_{\rm ap} M_{\times}^3\rangle$}}
\newcommand{\mxqd}{\mbox{$\langle M_{\times}^4\rangle$}}
\newcommand{\mqd}{\mbox{$\langle M^4\rangle$}}
\newcommand{\mcbmc}{\mbox{$\langle M^3 M^*\rangle$}}
\newcommand{\msqmcsq}{\mbox{$\langle M^2 M^{*2}\rangle$}}
\newcommand{\bfr}{\bmath{r}}
\newcommand{\bfk}{\bmath{k}}
\newcommand{\bfs}{\bmath{s}}
\newcommand{\bft}{\bmath{t}}
\newcommand{\bfq}{\bmath{q}}
\newcommand{\bfu}{\bmath{u}}
\newcommand{\bfx}{\bmath{x}}

\begin{document}

\maketitle

\begin{abstract}
We present simple formulae for calculating the skewness and 
kurtosis of the aperture mass statistic for weak lensing surveys 
which is insensitive to masking effects of survey geometry or variable
survey depth.  The calculations are the higher order analogs
of the formula given by \citet*{Sch02} which has been used to compute
the variance of the aperture mass from several lensing surveys.
As our formula requires
the three-point shear correlation function, we also 
present an efficient tree-based algorithm for measuring it. 
We show how our algorithm would scale in computing time and 
memory usage for future lensing surveys.
We also apply the procedure to our CTIO survey data, originally
described in \citet{Ja03}.  We find that the skewness is 
positive (inconsistent with zero) at the $2\sigma$ level.  
However, the signal is 
too noisy from this data to usefully constrain cosmology.
\end{abstract}

\begin{keywords}
gravitational lensing -- cosmology
\end{keywords}

\section{Introduction}
\label{introduction}

One of the most useful statistics for probing the power spectrum with
weak gravitational lensing is the aperture mass statistic, $\map$.
The aperture mass was first proposed by \citet*{Ka95} and \citet*{Sch96}
for estimating the masses of clusters, and
is essentially an estimate of the 
convergence, $\kappa$, within a circular aperture.  The
convergence is proportional to the projected mass
density relative to the average value in the field, hence
the name aperture mass.

Since the aperture mass is relative to the average value, the 
statistic $\langle\map\rangle$ gives 0 over the whole field.
However, higher order moments of the aperture mass are non-zero, 
and the variance,
$\mapsq$, has proved very useful for weak lensing researchers,
with many studies to date including this statistic in their
analysis \citep*[e.g.][]{Ho02, vW02, Ja03, Ha03}.  This statistic 
is a good probe of the power spectrum, and thus measures the
extent of the Gaussian fluctuations of the density of the universe.

However, any non-Gaussian component is not measured by the 
variance.  Further, the density fluctuations must be non-Gaussian,
since the density contrast is constrained to be greater than
-1, but it can be arbitrarily large in the positive direction.
Clusters generally have density contrasts of more than 200.  
Thus, the density fluctuations, which have a mean of 0 by 
definition, must be skewed towards positive values.
Indeed, this non-Gaussianity has recently been detected
in the VIRMOS-DESCART survey by two groups \citep*{Be02, Pen03}.

The lowest order measures of the non-Gaussianity using the 
shear field are called three-point statistics, 
since they require measurements of three shear values and their 
relative positions and orientations. 
Three-point statistics have the potential to be 
very useful for cosmology, since they can, in combination with
two-point statistics, determine $\Omega_m$ and 
$\sigma_8$ independently \citep*{Be97, Ja97, Sch98}.  
They can also be used to 
constrain the properties of halos such as the inner slope and
typical concentration parameters \citep*{Ta03}.
In this paper, we investigate a particular three-point 
statistic, the skewness of the aperture mass, $\mapcb$.

When \citet{Sch98} proposed the use of the aperture mass 
for cosmic shear measurements, they introduced the
form of the statistic which has been used by almost all 
weak lensing studies to date.  However, we find that calculations
of the three-point statistic is significantly easier using the 
form introduced by \citet*{Cr02}.  Namely,
\begin{align}
\label{mapdefu}
\map(R) &= \int d^2 \bfr U_R(r) \kappa(\bfr) \\
\label{mapdefq}
&= \int d^2 \bfr Q_R(r) \gamma_t(\bfr) \\
U_R(r) &= \frac{1}{2\pi R^2} 
  \left(1 - \frac{r^2}{2R^2}\right)
  \exp \left(-\frac{r^2}{2R^2}\right) \\
Q_R(r) 
&= -U_R(r) + \frac{2}{r^2} \int_0^r r' dr' U_R(r') \\
&= \frac{r^2}{4\pi R^4} 
  \exp \left(-\frac{r^2}{2R^2}\right) 
\end{align}
where $\kappa$ is the convergence, $\gamma_t$ is the tangential
component of the shear,
and $\bfr$ is measured from the centre of the aperture.

We also note that \citet*{Zh03} has shown that this form of the 
aperture mass is more sensitive for constraining $\Omega_m$ than other forms.

There are several features of the aperture mass statistic which makes it 
very useful.  First, it probes the power spectrum with a very 
narrow window function, $W(\eta)$.  \citet{Sch98} derive the relation:
\begin{align}
\mapsq(R) &= \frac{1}{2\pi} \int d^2 \bfk P(k) W(k R) \\
W(k R) &= \tilde U_R(k)^2 \\
&= \left[\frac{k^2 R^2}{2} \exp\left(-\frac{k^2 R^2}{2}\right)\right]^2 \\
W(\eta) &= \frac{\eta^4}{4} \exp(-\eta^2)
\end{align}
where the tilde ($~\tilde{}~$) indicates the Fourier transform.
This window function $W(\eta) \propto \eta^4$ for small $\eta$, and
drops super-exponentially for large $\eta$, peaking at $\eta = 2$.

Another benefit of the aperture mass is that it has (nearly) finite support
in real space, so it is calculable.  The ideal window function
would be a delta function, $W(\eta) = \delta_D(\eta-\eta_0)$, so
the statistic would directly probe $P(\eta_0/R)$.
However, to have infinitesimal extent in k-space would require
infinite extent in real space, and thus be incalculable.  \mapsq\
thus has the advantage that it is compact in real space as 
well\footnote{
Technically, the aperture mass does have infinite extent as well;
however, the exponential cutoff is so sharp that the window function
is effectively finite for real calculations.
}.

The third benefit is that it measures purely the so-called E-mode 
of the shear field.  There is a corresponding statistic, \mx, 
which measures the B-mode:
\begin{equation}
\label{mxdef}
\mx(R) = \int d^2 \bfr Q_R(x) \gamma_\times(\bfr) , 
\end{equation}
where the cross-component of the shear, $\gamma_\times$, is oriented
at an angle of $45$ degrees relative to $\gamma_t$. 
Then the variance of this measure, \mxsq\, is a measure of the B-mode power 
in the shear field, which is generally taken to be a measure of 
the contamination from residual systematics, such as 
uncorrected effects of the point-spread functions.  
Intrinsic alignments \citep{Cr02} and source clustering \citep{Sch02} can 
also produce B-mode signal, but the level of both of these is 
generall expected to be very low for scales larger than $1\arcmin$.

The only problem with calculating these statistics directly is that real
lensing surveys have regions which are masked out due to survey 
geometry, bright stars, bad seeing, etc.  Thus the aperture mass
statistic runs the risk of losing azimuthal symmetry due to the masking.
Since the statistic depends on azimuthal symmetry for the angular
integrals, a direct calculation of \mapsq\ and \mxsq\ will leak some 
of the E and B-mode power into the other statistic.  This may only be
of order a 10 per cent effect or less for typical surveys, but with the goal 
of precision cosmology, another method of calculation is typically 
used, first derived by \citet{Cr02} and then refined by \citet{Sch02}.  

They have shown that the aperture mass variance
can be calculated from an integral of the correlation functions in 
an equation of the form:
\begin{align}
\label{mapsqresult1}
\mapsq(R) &= \frac{1}{2} \int \frac{s ds}{R^2} \left[
  \xi_+(s) T_+\left(\frac{s}{R}\right) +
  \xi_-(s) T_-\left(\frac{s}{R}\right) \right] \\ 
\label{mxsqresult1}
\mxsq(R) &= \frac{1}{2} \int \frac{s ds}{R^2} \left[
  \xi_+(s) T_+\left(\frac{s}{R}\right) -
  \xi_-(s) T_-\left(\frac{s}{R}\right) \right]
\end{align}
where $\xi_{+,-}$ are the two-point shear correlation functions 
(defined more precisely in \S\ref{2ptsection})
and $T_{+,-}$ are known functions (also defined in \S\ref{2ptsection}).

The third moment of the aperture mass, \mapcb\ was originally suggested by 
\citet{Sch98} as a statistic for investigating the non-Gaussianity 
of the shear field.  It is useful as a cosmological probe, since 
(as these authors showed) its
dependence on $\Omega_m$ and $\sigma_8$ are somewhat orthogonal to that
of \mapsq, so the combination of the two statistics
can be used to determine $\Omega_m$ and $\sigma_8$ separately.
In particular $\mapcb/\mapsq^2$ scales approximately as $1/\Omega_m$,
independent of $\sigma_8$.

However, a direct measurement of \mapcb\ using apertures
would be even more affected by masking than the variance is.  
Thus, in this paper, we derive formulae similar to 
Equations~(\ref{mapsqresult1},\ref{mxsqresult1})
for \mapcb\ and \mxcb\  in terms of the three-point correlation function.

The signal-to-noise (S/N) for \mapcb\ is significantly lower than for \mapsq.
As we will show below, our total S/N for \mapcb\ is of 
order unity for the roughly $10^6$ galaxies in our CTIO survey, 
compared to a S/N of 
about 7 for \mapsq\ \citep{Ja03}.  As the S/N scales as $N_{\rm gal}^{1/2}$,
we would need about $10^8$ galaxies to make a good measurement of 
\mapcb.  Planned surveys such as those from the Supernova
Anisotropy Probe (SNAP) and the Large-aperture Synoptic Survey
Telescope (LSST) will do just that.  Actually, the S/N also increases when one 
goes deeper, which both of these telescopes will do, so their S/N
should be somewhat better than 10.
In any case, the large number of galaxies from these future missions
demands efficient algorithms for calculating \mapcb,
such as the one presented herein.

In \S\ref{2ptsection}, we define our notations and 
briefly rederive the above equations (\ref{mapsqresult1},\ref{mxsqresult1})
in a manner that will lend
itself to generalization to the three-point version. 
This is important, since we found that the derivations given
by \citet{Cr02} and \citet{Sch02} do not generalize easily 
to the three-point case.  The method is fairly similar to that of
\citet{Pen03}, although they use a tensor formulation, 
which is a bit unwieldy, having 64 components for the three-point
correlation function (8 of which are unique).
Then \S\ref{3ptsection}
will derive the corresponding equations for the three-point statistic.
\S\ref{algorithmsection} describes our algorithm for calculating the 
three-point correlation function.
In \S\ref{applicationsection}, we then apply the formulae to simulated data 
to check their validity, and also to our CTIO survey data.

\section{Variance of the Aperture Mass}
\label{2ptsection}

We follow the notation of \citet{Sch02} to describe the E and B-mode 
components of the shear field by defining the lensing potential, 
$\psi$, to be complex.  
\begin{equation}
\psi = \psi^E + i \psi^B
\end{equation}

The convergence, $\kappa$, and shear, $\gamma$, are related to the potential by
\begin{align}
\kappa &= \frac{1}{2} \nabla^2 \psi \\
\gamma &= \frac{1}{2} \left(\frac{\partial}{\partial x} 
                           + i \frac{\partial}{\partial y}\right)^2 \psi
\end{align}
Since the convergence is the projected matter density, it is real for 
pure lensing fields.  Thus, lensing produces only E-mode fields,
and the B-mode is generally used to check for residual systematics
or the aforementioned effects of intrinsic alignments or sourse clustering.

With a complex $\kappa$, \eqq{mapdefu} would give us a complex value
for $\map$.  However, using \eqq{mxdef}, and defining 
$M = \map + i \mx$, we find
\begin{align}
\label{mkappa}
M(R) &= \int d^2 \bfr U_R(r) \kappa(\bfr) \\
\label{mshear}
&= -\int d^2 \bfr Q_R(r) \gamma(\bfr) e^{-2i\phi}
\end{align}
where $\phi$ is the polar angle of $\bfr$. 

Using this definition, the expectation value of $M^2$ is
\begin{equation}
\msq(R) 
= \int d^2 \bfr_1 \int d^2 \bfr_2 Q_R(r_1) Q_R(r_2) 
   \langle\gamma(\bfr_1)\gamma(\bfr_2)\rangle 
   e^{-2i(\phi_1 + \phi_2)}
\label{eqa}
\end{equation}

We now define the `natural components' of the two-point shear 
correlation function as:
\begin{align}
\label{ximinusdef}
\xi_-(s) &= \langle \gamma(\bfr) \gamma(\bfr + \bfs) e^{-4i\alpha} \rangle \\
\label{xiplusdef}
\xi_+(s) &= \langle \gamma(\bfr) \gamma^*(\bfr + \bfs) \rangle 
\end{align}
where $\alpha$ is the polar angle of $\bfs$.  

Then, \eqq{eqa} becomes (taking $\bfr_2 = \bfr_1 + \bfs$, and 
dropping the subscript for $\bfr_1$)
\begin{equation}
\msq(R) 
= \int d^2 \bfr \int d^2 \bfs Q_R(r) Q_R(|\bfr+\bfs|) 
   \xi_-(s) e^{2i(2\alpha-\phi_1-\phi_2)}
\label{eqb}
\end{equation}

To evaluate this, we treat $\bfr$ and $\bfs$ as complex numbers,
so $\bfs = s \exp(i\alpha)$, $\bfr = r \exp(i\phi_1)$, and 
$\bfr + \bfs = |\bfr + \bfs| \exp(i\phi_2)$.  Then
\begin{align}
\label{mm_a}
\msq(R) 
&= \int d^2 \bfs \xi_-(s) \frac{\bfs^4}{s^4} 
    \int d^2 \bfr Q_R(r_1) Q_R(|\bfr+\bfs|)
    \frac{(\bfr^*+\bfs^*)^2}{|\bfr+\bfs|^2}
    \frac{\bfr^{*2}}{r^2} \\
&= \int \frac{s ds}{R^2} \xi_-(s) \left[\frac{s^4}{128 R^4} 
    \exp\left(-\frac{s^2}{4R^2}\right) \right] \\
&\equiv \int \frac{s ds}{R^2} \xi_-(s) T_-\left(\frac{s}{R}\right)
\end{align}
where ${}^*$ indicates complex conjugate.  (The evaluation
of the integral in \eqq{mm_a} is straightforward in Cartesian coordinates, 
where each complex value is represented as, for example,
$\bfr = x + iy$.)

A similar procedure can be used to evaluate \mmc:
\begin{align}
\mmc(R)
&= \int d^2 \bfr \int d^2 \bfs Q_R(r) Q_R(|\bfr+\bfs|) 
   \langle\gamma(\bfr)\gamma^*(\bfr+\bfs)\rangle 
   e^{-2i(\phi_1 - \phi_2)} \\
&= \int d^2 \bfs \xi_+(s) 
    \int d^2 \bfr Q_R(r) Q_R(|\bfr+\bfs|)
    \frac{(\bfr+\bfs)^2}{|\bfr+\bfs|^2}
    \frac{\bfr^{*2}}{r^2} \\
&= \int \frac{s ds}{R^2} \xi_+(s) 
    \left[\frac{s^4-16R^2s^2+32R^4}{128 R^4} 
    \exp\left(-\frac{s^2}{4R^2}\right) \right] \\
&\equiv \int \frac{s ds}{R^2} \xi_+(s) T_+\left(\frac{s}{R}\right)
\end{align}

Finally we can convert these expressions into formulae for \mapsq\
and \mxsq\ individually through the expressions:
\begin{align}
\mapsq + i \mapmx &= \frac{1}{2} \left(\mmc + \msq \right) \\
\mxsq - i \mapmx &= \frac{1}{2} \left(\mmc - \msq \right)
\end{align}
So,
\begin{align}
\label{mapsqform}
\mapsq(R) + i\mapmx(R)
&= \frac{1}{2} \int s ds
   \left[ \xi_+(s) T_+\left(\frac{s}{R}\right) +
          \xi_-(s) T_-\left(\frac{s}{R}\right) \right] \\
\label{mxsqform}
\mxsq(R) - i\mapmx(R) 
&= \frac{1}{2} \int s ds
   \left[ \xi_+(s) T_+\left(\frac{s}{R}\right) -
          \xi_-(s) T_-\left(\frac{s}{R}\right) \right]
\end{align}
where
\begin{align}
T_+(x) 
&= \frac{x^4-16x^2+32}{128} 
    \exp\left(-\frac{x^2}{4}\right) \\
T_-(x)
&= \frac{x^4}{128} 
    \exp\left(-\frac{x^2}{4}\right) 
\end{align}

This result agrees with that obtained from the formulae in 
\citet{Sch02} as well as the formulae in 
\citet{Cr02}\footnote{ 
We note however, that the results quoted in \citet{Cr02} for $\cal W$ 
and $\tilde{\cal W}$ 
(which correspond to our $T_+$ and $T_-$, 
but for a slightly different normalization for $U$) are too low by a 
factor of $\pi/2$.
}. 
Also, \citet{Pen03} gives a formula for a tensor form of
Equations~\ref{mapsqform} and \ref{mxsqform}, which can be 
expanded to agree with our results.

\section{Third Moment of the Aperture Mass}
\label{3ptsection}

We now follow the same procedure as in \S\ref{2ptsection} for evaluating
\mcb\ and \msqmc.  

\begin{equation}
\mcb(R) 
= -\int d^2\bfr \int d^2\bfs \int d^2\bft
   Q_R(r) Q_R(|\bfr+\bfs|) Q_R(|\bfr+\bft|) 
   \langle\gamma(\bfr)\gamma(\bfr+\bfs)\gamma(\bfr+\bft)\rangle 
   e^{-2i(\phi_1 + \phi_2 + \phi_3)} 
\label{eqc}
\end{equation}

\begin{figure}
\includegraphics[width=60mm]{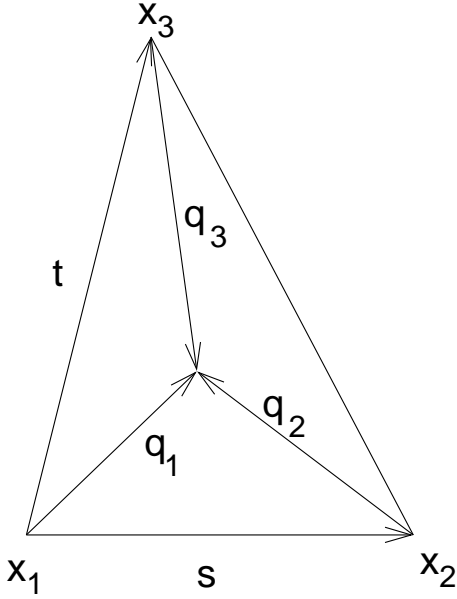}
\caption[]{\small
Graphical representation of the $q$ parameters used in the formulae for $T_0$
and $T_1$.  These vectors are used as complex numbers in the formulae.}
\label{qfig}
\end{figure}

For the `natural components' of the three-point correlation function, 
we follow \citet*{SL} and define:
\begin{align}
\Gamma_0(s,\bft^\prime) 
&= \langle \gamma(\bfr) \gamma(\bfr+\bfs) \gamma(\bfr+\bft) 
   e^{-2i(\alpha_1+\alpha_2+\alpha_3)} \rangle \\
\Gamma_1(s,\bft^\prime) 
&= \langle \gamma^*(\bfr) \gamma(\bfr+\bfs) \gamma(\bfr+\bft) 
   e^{-2i(-\alpha_1+\alpha_2+\alpha_3)} \rangle \\
\Gamma_2(s,\bft^\prime) 
&= \langle \gamma(\bfr) \gamma^*(\bfr+\bfs) \gamma(\bfr+\bft) 
   e^{-2i(\alpha_1-\alpha_2+\alpha_3)} \rangle \\
\Gamma_3(s,\bft^\prime) 
&= \langle \gamma(\bfr) \gamma(\bfr+\bfs) \gamma^*(\bfr+\bft) 
   e^{-2i(\alpha_1+\alpha_2-\alpha_3)} \rangle 
\end{align}
where $\bft^\prime = \bft \bfs^*/s$.  Since we have three points now, 
they form a triangle, which is depicted in Fig.~\ref{qfig}.
The parameter $\bft^\prime$ is simply $\bft$ when the triangle is oriented such that
$\bfs$ is parallel to the x-axis.  Since
the correlation functions are rotationally invariant, they are 
properly a function of only three real variables, such as $(s,\bft^\prime)$, 
rather than four $(\bfs,\bft)$.

\citet{SL} give several possible definitions for $\alpha_i$. 
For our purposes, it does not much matter which definition one 
takes, as long as the $\alpha$'s correspond to directions
which rotate with the triangle.
The simplest example is for all three to be equal to the 
polar angle of $\bfs$\footnote{
For other definitions of these angles, $\alpha_i^\prime$, 
simply multiply the final formula for $T_0$ by 
$\exp(2i(\alpha_1^\prime + \alpha_2^\prime + \alpha_3^\prime) - 6i\alpha)$
and $T_1$ by 
$\exp(2i(-\alpha_1^\prime + \alpha_2^\prime + \alpha_3^\prime) - 2i\alpha)$.
}.
\begin{align}
\bfs &= s e^{i \alpha} \\
\alpha_1 &= \alpha_2 = \alpha_3 = \alpha
\end{align}

Then, with these definitions,
\begin{align}
\mcb(R)
&= -\int d^2\bfr \int d^2\bfs \int d^2\bft
   Q_R(r) Q_R(|\bfr+\bfs|) Q_R(|\bfr+\bft|) 
   \Gamma_0(s,\bft^\prime) 
   e^{6i\alpha - 2i(\phi_1 + \phi_2 + \phi_3)} \\
\label{mmminta}
&= -\int d^2\bfs \int d^2\bft 
   \Gamma_0(s,\bft^\prime) \frac{\bfs^6}{s^6} 
   \int d^2\bfr Q_R(r) Q_R(|\bfr+\bfs|) Q_R(|\bfr+\bft|) 
   \frac{\bfr^{*2} (\bfr^*+\bfs^*)^2 (\bfr^*+\bft^*)^2}{
         r^2 |\bfr+\bfs|^2 |\bfr+\bft|^2} \\
\label{mmmintb}
&= \int \frac{s ds}{R^2} \int \frac{d^2\bft^\prime}{2 \pi R^2} 
   \Gamma_0(s,\bft^\prime) \left[ -\frac{\bfq_1^{*2} \bfq_2^{*2} \bfq_3^{*2}}{24 R^6}
   \exp\left(-\frac{q_1^2+q_2^2+q_3^2}{2R^2}\right) \right] \\
&\equiv \int \frac{s ds}{R^2} \int \frac{d^2\bft^\prime}{2 \pi R^2} 
   \Gamma_0(s,\bft^\prime) T_0\left(\frac{s}{R},\frac{\bft^\prime}{R}\right)
\label{mmmint}
\end{align}
where $\bfq_i$ are the vectors from each vertex to the 
centroid of the triangle\footnote{
These are similar to the $q_i$ vectors used by \citet{Pen03}.  
Their $q_i$'s are 3 times the values our $q_i$'s.  
}.  
They are depicted graphically in Fig.~\ref{qfig}.  
Algebraically, 
\begin{align}
\bfq_1 &= (s + \bft^\prime)/3 \\
\bfq_2 &= (\bft^\prime - 2s)/3 \\
\bfq_3 &= (s - 2\bft^\prime)/3 
\end{align}

Likewise,
\begin{align}
\nonumber
\msqmc(R)
&= -\int d^2\bfr \int d^2\bfs \int d^2\bft
   Q_R(r) Q_R(|\bfr+\bfs|) Q_R(|\bfr+\bft|) 
   \Gamma_1(s,\bft^\prime)
   e^{2i\alpha - 2i(-\phi_1 + \phi_2 + \phi_3)} \\
&= \int \frac{s ds}{R^2} \int \frac{d^2\bft^\prime}{2 \pi R^2} \Gamma_1(s,\bft^\prime) 
   \left[-
   \left(\frac{\bfq_1^2 \bfq_2^{*2} \bfq_3^{*2}}{24 R^6} 
         - \frac{q_1^2 \bfq_2^* \bfq_3^*}{9 R^4}
         + \frac{\bfq_1^{*2} + 2\bfq_2^* \bfq_3^*}{27 R^2} \right)
   \exp\left(-\frac{q_1^2+q_2^2+q_3^2}{2R^2}\right) 
   \right] \\
&\equiv \int \frac{s ds}{R^2} \int \frac{d^2\bft^\prime}{2 \pi R^2} 
   \Gamma_1(s,\bft^\prime) T_1\left(\frac{s}{R},\frac{\bft^\prime}{R}\right)
\label{mmmcint}
\end{align}

The functions $T_0$ and $T_1$ are thus:
\begin{align}
\label{t0eqn}
T_0(s,\bft) &= -\frac{\bfq_1^{*2} \bfq_2^{*2} \bfq_3^{*2}}{24}
   \exp\left(-\frac{q_1^2+q_2^2+q_3^2}{2}\right) \\
T_1(s,\bft) &= 
   -\left( \frac{\bfq_1^2 \bfq_2^{*2} \bfq_3^{*2}}{24}
   - \frac{q_1^2 \bfq_2^* \bfq_3^* }{9}
   + \frac{\bfq_1^{*2} + 2\bfq_2^* \bfq_3^*}{27} \right)
   \exp\left(-\frac{q_1^2+q_2^2+q_3^2}{2}\right) 
\label{t1eqn}
\end{align}
where $s$, $\bft$, and the $\bfq$'s are now dimensionless quantities.

As an aside, we note that integral formulations for $T_0$ and $T_1$ 
are possible for any aperture function $U_R(r)$.  However, the 
particular form we use here is the only one for which we have been
able to calculate closed forms for $T_0$ and $T_1$.  In particular,
the Gaussian exponential term makes the integration step from \eqq{mmminta} to
(\ref{mmmintb}) (and the corresponding step for $T_1$) straightforward,
although rather messy.

\begin{figure}
\centering
\includegraphics[height=100mm,angle=-90]{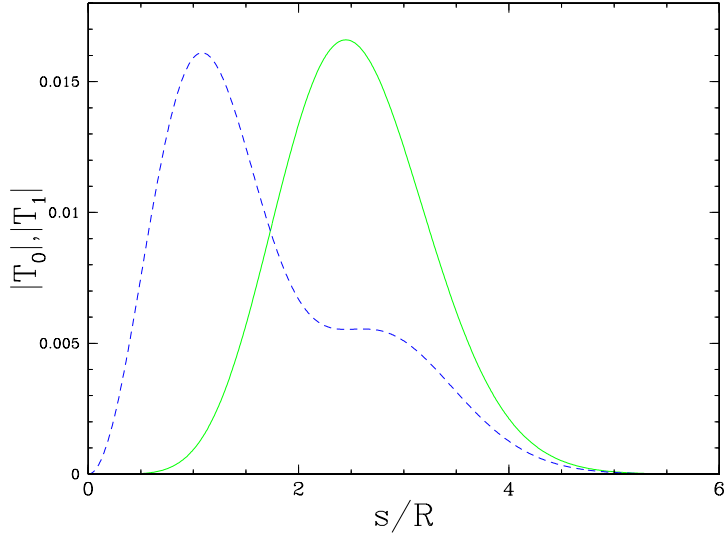}
\caption[]{\small
The absolute magnitude of the functions $T_0$ and $T_1$ 
for equilateral 
triangles as a function of the side length of the triangle, $s$.
The solid green curve is $|T_0|$, and the dashed blue curve is
$|T_1|$.
}
\label{t0t1figure}
\end{figure}

Fig.~\ref{t0t1figure} shows the absolute value of $T_0$ and $T_1$
for equilateral triangles as a function of the triangle side length.
Note that the $T_i$ are significant for triangles with side lengths
up to about $4R$, so to measure $\mapcb(R)$ up to $R = 10\arcmin$, one would
need a survey at least about $40\arcmin$ on a side.  
Interestingly, this is about the same size as one needs for the 
corresponding measurement of $\mapsq(R)$.

Finally, we use the relations:
\begin{align}
\mcb &= \mapcb +3i \mapsqmx -3 \mapmxsq -i \mxcb \\
\msqmc &= \mapcb +i\mapsqmx +\mapmxsq +i\mxcb
\end{align}
to obtain:
\begin{align}
\label{mapcbint}
\mapcb(R)   &= \frac{1}{4} \Re\left(3 \msqmc(R) + \mcb(R) \right) \\
\mapsqmx(R) &= \frac{1}{4} \Im\left(\msqmc(R) + \mcb(R) \right) \\
\mapmxsq(R) &= \frac{1}{4} \Re\left(\msqmc(R) - \mcb(R) \right) \\
\mxcb(R)    &= \frac{1}{4} \Im\left(3 \msqmc(R) - \mcb(R) \right)
\end{align}

There is one further complication for using these formulae.
In practice, one does not calculate $\Gamma_i(s,\bft)$ for every $(s,\bft)$,
since this would include every triangle six times.  A typical
program would bin each triangle according to its shortest 
side, s, and its second shortest side, t, thus counting every triangle only once.

In this case, let us assume that we know $\Gamma_i(s,\bft)$ for all
$s$, but for only those $\bft$ where $|\bft~-~s|~>~t~>~s$.
Then naively, \eqq{mmmint} becomes
\begin{align}
\nonumber
\mcb(R) = 
& \left[\int \frac{s ds}{R^2}\int\limits_{s<t^\prime<|\bft^\prime-s|} \frac{d^2\bft^\prime}{2\pi R^2} 
   \Gamma_0(s,\bft^\prime) 
   T_0\left(\frac{s}{R},\frac{\bft^\prime}{R}\right) \right.  \\ 
\nonumber 
& + \int \frac{s ds}{R^2} \int\limits_{s<|\bft^\prime-s|<t^\prime} \frac{d^2\bft^\prime}{2\pi R^2} 
   \Gamma_0(s,s-\bft^\prime) 
   T_0\left(\frac{s}{R},\frac{s-\bft^\prime}{R}\right) \\
\nonumber
& + \int \frac{s ds}{R^2}\int\limits_{t^\prime<s<|\bft^\prime-s|} \frac{d^2\bft^\prime}{2\pi R^2} 
   \Gamma_0\left(t^\prime,\frac{s\bft^{\prime*}}{t^\prime}\right) 
   T_0\left(\frac{t^\prime}{R},\frac{s\bft^{\prime*}}{t^\prime R}\right) \\
\nonumber 
& + \int \frac{s ds}{R^2}\int\limits_{t^\prime<|\bft^\prime-s|<s} \frac{d^2\bft^\prime}{2\pi R^2} 
   \Gamma_0\left(t^\prime,\frac{(\bft^\prime-s)\bft^{\prime*}}{t^\prime}\right) 
   T_0\left(\frac{t^\prime}{R},\frac{(\bft^\prime-s)\bft^{\prime*}}{t^\prime R}\right) \\
\nonumber
& + \int \frac{s ds}{R^2}\int\limits_{|\bft^\prime-s|<s<t^\prime} \frac{d^2\bft^\prime}{2\pi R^2} 
   \Gamma_0\left(|\bft^\prime-s|,\frac{-s(\bft^\prime-s)^*}{|\bft^\prime-s|}\right) 
   T_0\left(\frac{|\bft^\prime-s|}{R},\frac{-s(\bft^\prime-s)^*}{|\bft^\prime-s| R}\right) \\
& \left.+ \int \frac{s ds}{R^2}\int\limits_{|\bft^\prime-s|<t^\prime<s} \frac{d^2\bft^\prime}{2\pi R^2} 
   \Gamma_0\left(|\bft^\prime-s|,\frac{\bft^\prime(\bft^\prime-s)^*}{|\bft^\prime-s|}\right) 
   T_0\left(\frac{|\bft^\prime-s|}{R},\frac{\bft^\prime(\bft^\prime-s)^*}{|\bft^\prime-s| R}\right)  \right]
\end{align}

However, for each of these integrals, a change of variables
transforms it into the first one.  Thus, we have the much simpler
result:
\begin{equation}
\mcb(R) = 
6 \int \frac{s ds}{R^2}\int\limits_{s<t^\prime<|\bft^\prime-s|} \frac{d^2\bft^\prime}{2\pi R^2} 
   \Gamma_0(s,\bft^\prime) T_0\left(\frac{s}{R},\frac{\bft^\prime}{R}\right)
\label{mcbform}
\end{equation}

The case is almost the same for \msqmc, except that the conjugated vertex
changes in 4 of the integrals, so the net result is:
\begin{align}
%\nonumber
\msqmc(R) &= 
2 \int \frac{s ds}{R^2}\int\limits_{s<t^\prime<|\bft^\prime-s|} \frac{d^2\bft^\prime}{2\pi R^2} 
\left[
   \Gamma_1(s,\bft^\prime) T_1\left(\frac{s}{R},\frac{\bft^\prime}{R}\right) %\right. \\
%&\qquad \left.
   + \Gamma_2(s,\bft^\prime) T_2\left(\frac{s}{R},\frac{\bft^\prime}{R}\right)
   + \Gamma_3(s,\bft^\prime) T_3\left(\frac{s}{R},\frac{\bft^\prime}{R}\right) \right]
\label{msqmcform}
\end{align}
where $T_2$ and $T_3$ are obtained from $T_1$ by a cyclic rotation of 
the indices.

The extension to the kurtosis of the aperture mass 
($\langle M_{\rm ap}^4 \rangle(R)$)
is straightforward and is given in the appendix.

\section{Efficient Calculation of the Three-Point Correlation Function}
\label{algorithmsection}

\subsection{Two-Point Algorithm}

We consider an efficient algorithm for calculating the two-point
correlation function first.  This problem has been solved in a number
of ways for the purposes of number count correlations for studying 
large-scale structure.  \citet*{Be98} gives a review of some techniques 
for that purpose.  The algorithm we present below is most similar
to that of \citet*{Mo00}.  There are a few complications due to the 
fact that we are correlating a vector field rather than a scalar 
number count, but the creation and use of the cells are quite
analagous.  Also, \citet*{Zh04} have recently presented a similar 
algorithm for the purposes of weak lensing.
Therefore, we present a brief description of the algorithm below
in order to describe the unique features of our version
of the algorithms, and we 
defer to these other two papers for more comprehensive descriptions of the 
general method.

The most naive algorithm for calculating the two-point correlation
function would be to take every pair of galaxies in the 
field and put the product of the shears into a bin corresponding to their
separation.  Then the average value could be found
for each bin.  Clearly, this is extremely slow for large datasets, being
an $O(N^2)$ algorithm.  The corresponding three-point algorithm is even 
worse, $O(N^3)$.

However, we note that the binning inherent in even this brute-force
technique is effectively smoothing the correlation function by the 
size of the bins, since we lose all information about where {\em within}
each bin the pair really belongs.  If we allow ourselves to smooth 
again on this scale, then we can greatly speed up the algorithm.

First, binning is usually done in logarithmic bins in the separation, $d$,
with some specified minimum and maximum scale, $d_{min} < d < d_{max}$.
Now, suppose we have two circular regions of radii $s_1$ and $s_2$,
whose centres are separated by a distance $d_c$.  Then, 
then the separation for any pair of galaxies taken from these two regions
must fall in the range $d_c - s_1 - s_2 < d < d_c + s_1 + s_2$.  If the 
radii are much less than $d$, this becomes
\begin{equation}
\ln(d_c) - \frac{s_1 + s_2}{d_c} < \ln(d) < \ln(d_c) + \frac{s_1 + s_2}{d_c}
\end{equation}
So, if our regions satisfy
\begin{equation}
\frac{s_1 + s_2}{d_c} < b
\label{soverd}
\end{equation}
where $b\equiv \Delta ({\rm ln} d)$ 
is the bin size, then we can put all of them into the same bin, 
knowing that we will be wrong by at most one bin for any specific pair 
of galaxies.

In this case, we can take the average of the shear values in each region
and place the product of the averages in our bin, dropping the calculation
time for these galaxies from $O(N_1N_2)$ to $O(N_1+N_2)$, where $N_1$ and 
$N_2$ are the number of galaxies in each region.  When the $N$'s 
are large, this can be a huge speedup.

The specific technique for implementing this type of algorithm is to 
create a so-called kd-tree (with k=2 in this case) for the data.
Again, we defer to \cite{Mo00} and \citet{Zh04} for a more complete 
description of this type of data structure.

In short, we bin the data in `Cells', keeping track of various average
values for the galaxies in each Cell along with its centroid and 
size (maximum distance of a galaxy from the centroid).  
Each Cell with more than one galaxy (and larger than $d_{\rm min}$) 
also contains pointers to two 
sub-Cells with half the number of galaxies each.  

When \eqq{soverd} is not satisfied, we split each Cell (roughly) in 
half and check the resulting pairs of regions.  
Cell splitting continues in 
this manner until all pairs can be calculated directly.  Note that the 
recursion must stop at some point, 
since $s_1 + s_2 = 0$ for two individual galaxies, 
so \eqq{soverd} must be satisfied eventually.

\begin{figure}
\includegraphics[width=120mm]{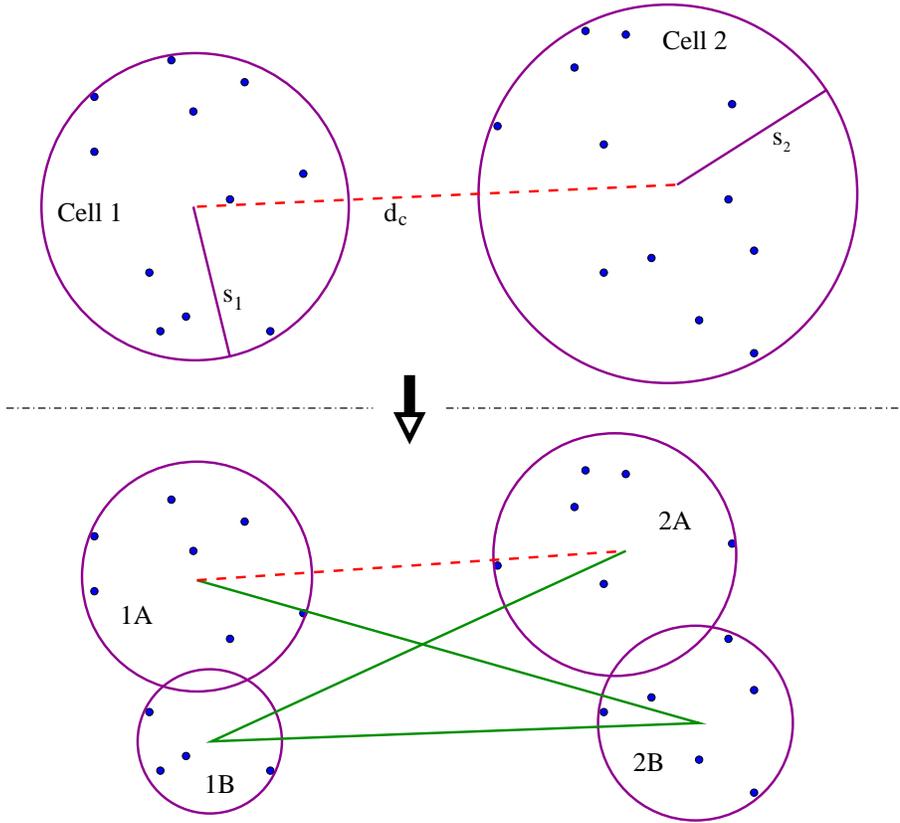}
\caption[]{\small
A sample calculation step in the two-point algorithm.  At first
the two Cells are too large compared to the distance between
them, so they need to be split up.  After the splits, there are
four pairs of subcells to be considered.  Three of the pairs 
(marked with green solid lines)
satisfy \eqq{soverd}, and thus can be computed directly.
The fourth (marked with a red dashed line) needs to be split again.
}
\label{splitfigure}
\end{figure}

Fig.~\ref{splitfigure} shows an example of a splitting decision in 
the algorithm.  At this point we are calculating the correlation between 
all the points in Cell 1 with all the points in Cell 2.  In the top half
of the figure, $s_1 + s_2$ is too large compared to $d_c$, so we
split both Cells into their subcells.  There are now 4 pairs of subcells to 
consider.  The three marked by the green solid line now satisfy
\eqq{soverd}, and can be calculated directly.  
But the 1A-2A pair marked by the
red dashed line needs to be split again.  

We tested this algorithm on a 
2.4 GHz Xeon processor for a square field 
200 arcminutes on a side, and found the calculation time and memory
usage to be\footnote{
Technically, there must also be
a $\ln(N)$ factor in the memory, and possibly the time, 
but this factor is essentially a constant over the range 
of $N$ we tested, so we neglect it.
}:
\begin{align}
T &\approx 3.6
  \left(\frac{N}{10^5}\right)^{1.0} 
  \left(\frac{b}{0.1}\right)^{-1.7} \text{seconds} \\
M &\approx 19
  \left(\frac{N}{10^5}\right)^{1.0} \text{MB}
\end{align}
Even for $N \approx 10^8$, which will be required for surveys like those of
the Supernova Anisotropy Probe (SNAP) and the Large-aperture Synoptic
Survey Telescope (LSST), this algorithm still only takes about an hour and uses less
than 20 GB of memory, which we expect will be common on destop 
computers by the time such surveys become available.
Therefore, we do not foresee any need to improve upon this algorithm in the 
near future.

\subsection{Three-Point Algorithm}

For the three-point function, the triangles are parametrized by three
values (since we ignore any overall rotation or translation).
One possible parametrization would be to
use the three side lengths: $d_1 > d_2 > d_3$.  This is not a good choice for
two reasons.  First, the orientation of the points is lost - 
a non-isocolese triangle is not distinguished from its mirror image.
Second, the range of the parameters is a function of the other 
parameter values (e.g. $d_1-d_2 < d_3 < d_2$).

A better choice is the following for a triangle with $d_1 > d_2 > d_3$: 
\begin{alignat}{2}
r &= d_3 & \qquad r_{\rm min} < r& < r_{\rm max} \\
u &= d_3/d_2 & 0 < u& < 1\\
v &= \pm (d_1-d_2)/d_3 & -1 < v& < 1
\end{alignat}
where v is positive when $(\bfx_1,\bfx_2,\bfx_3)$ are oriented counter-clockwise 
and negative when clockwise.

\begin{figure}
\includegraphics[width=120mm]{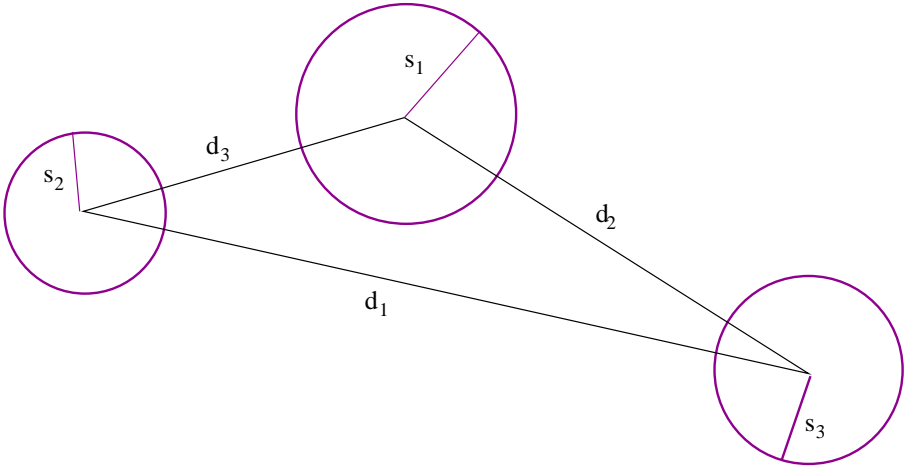}
\caption[]{\small
The geometry of the triangle in the discussion of the three-point 
algorithm.  We use the slightly non-intuitive convention that $d_1 > d_2 > d_3$
(rather than the other way around).  This is so we agree with Fig.~\ref{qfig}
with s as the smallest side, as desired for Equations~\ref{mcbform} and
\ref{msqmcform}.
}
\label{trianglefigure}
\end{figure}

We bin uniformly in ln($r$), $u$ and $v$ with a bin size $b$.
Now, if we have three circular regions of radii $s_1, s_2, s_3$, 
whose centres make a triangle with $d_1 > d_2 > d_3$ 
(see Fig.~\ref{trianglefigure}),
we can use the average shear in each when all of the following conditions hold:
\begin{align}
\frac{s_1+s_2}{d_3} &< b \\
\frac{s_3}{d_2} &< \frac{b}{u} \\
\frac{s_3}{d_2} &< \frac{b}{\sqrt{1-v^2}}
\end{align}

When considering three Cells for calculating the correlation
function, if the sizes satisfy the 
above criteria, we use the average shears for each and place the 
product into the bin corresponding to the triangle formed by their
centres.  Otherwise, we
recurse down to the sub-Cells of (some or all of) the Cells and
try again.  

This algorithm is far faster than the naive brute-force approach
of taking every triple of galaxies which would be $O(N^3)$. 
For the same field and processor mentioned above, we find the computation
time and memory usage for our implementation of this algorithm to be:
\begin{align}
T &\approx 415
  \left(\frac{N}{10^5}\right)^{1.4} 
  \left(\frac{b}{0.1}\right)^{-3.3} \text{minutes} \\
M &\approx 22 
  \left(\frac{N}{10^5}\right)^{1.0} \text{MB}
\end{align}

\citet{Zh04} show some empirical computation times for their algorithm
as well, which are shown in their fig.~3.
Although their tests were made at somewhat lower values of $N$ than
our tests, we believe that their scaling law at the same fiducial 
values we quote above is likely to be similar to ours.

The algorithm takes the most time dealing with triangles which are a 
significant fraction of the size of the field.  So for very large
fields, when one is only interested in the correlation on small
scales, the calculation will be somewhat faster than this, as
the larger triangles can be ignored.

\subsection{A Faster Algorithm}

We have discovered a way to significantly improve upon this algorithm, 
at least for large values of $N$.  However, it can require a significant
amount of memory, so depending on one's computing resources and 
survey size, it may not always be a viable option.  

\begin{figure}
\includegraphics[width=120mm]{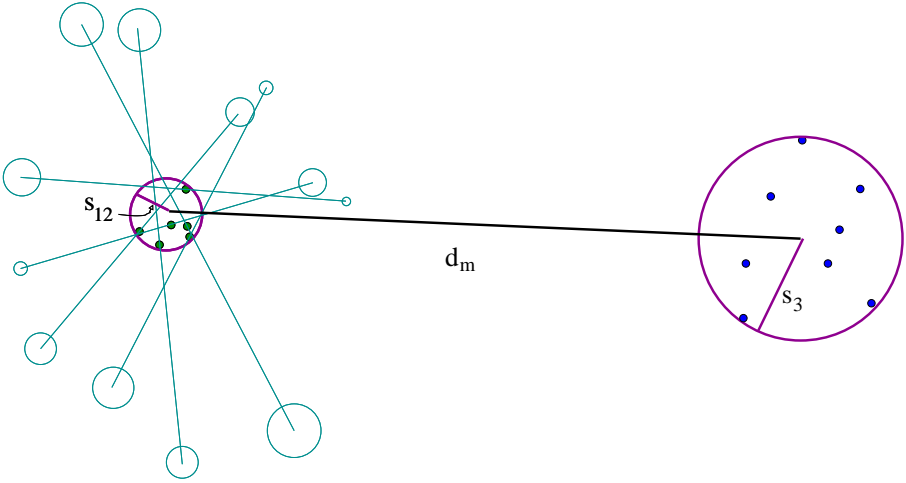}
\caption[]{\small
The geometry of the PairCell's for the improved three-point algorithm.
The pairs on the left side of the figure all have midpoints which are
near each other, and have cells at roughly the same separation.  Thus, 
they all fall into a single PairCell.  The dots represent the midpoints
of each pair.  The size of the PairCell,
$s_{12}$, is based on the scatter of these midpoints.  The distance from 
the PairCell to another Cell is then the median of the corresponding triangles,
$d_m$.
}
\label{paircellfigure}
\end{figure}

In this algorithm, we proceed one bin at a time in $\ln(r)$ rather than
doing all at once as we did in the above algorithm.  For each bin, we find
all pairs of Cells with $(s_1+s_2)/d_3 < b$ where $d_3$ falls in
the bin in question.  Then, we find the midpoint $M$ of each pair,
and create a new kd-tree for these pairs using these $M$'s
for the positions of each pair.
(See Fig.~\ref{paircellfigure}.)

Then, each `PairCell' refers to a collection of pairs.
For each, we keep track of the mean $M$ position, 
and the scatter of the $M$'s, which we call $s_{12}$. 
We don't average the data for all of the pairs in the bin, since the 
triangles made from these pairs and a third point can
still have very different shapes depending on the orientation 
of the pair.  But two pair 
with similar orientations will go into the same 
$r,u,v$ bin.  Therefore, within each PairCell, we bin the pairs
according to their orientation, using a total of 4/b bins
in order to maintain our rule of being off by at most one bin
in u and v.

Finally we correlate the PairCell data with 
the third points of the triangles by traversing the tree of 
pairs and the original tree of Cells described above.  For a 
given PairCell with size $s_{12}$ and a normal Cell with 
size $s_3$, separated by a distance $d_m$\footnote{The m stands
for median, since this line is the median of the triangle from 
vertex 3}, we split one or both unless
\begin{equation}
\frac{s_{12} + s_3}{d_m} < b
\label{pairsoverd}
\end{equation}

Note that for each PairCell and Cell which satisfy this requirement, 
the various pair orientations stored in the PairCell span the entire
range of $-1 < v < 1$.  Also, if $d_3/d_m$ is not very small, we need to
calculate $u$ separately for each orientation as well, since $u$
can vary by several bin sizes.  But for each 
individual orientation, the above equation guarantees that we are 
only smoothing by one bin in $u$ or $v$.

When one typically has many pairs in each PairCell at the point where
\eqq{pairsoverd} holds, then this algorithm would be expected to be
significantly faster than our previous algorithm, which effectively
does each pair individually.  This is usually true 
for the large triangles where the algorithm spends the most time,
so we find that this algorithm does turn out to be significantly
faster than our previous one.

Empirically, the calculation time and memory usage for this algorithm 
are found to be:
\begin{align}
T &\approx 161 
  \left(\frac{N}{10^5}\right)^{1.1} 
  \left(\frac{b}{0.1}\right)^{-3.7} \text{minutes} \\
M &\approx 800
  \left(\frac{N}{10^5}\right)^{1.0}
  \left(\frac{b}{0.1}\right)^{-1.2} \text{MB}
\end{align}
This is a factor of 2.6 faster than the previous algorithm for
our fiducial values, and the speedup increases with $N$, since it scales
as a smaller power of $N$ than our previous algorithm.

However, for fields with $N > 10^6$, the memory demands become difficult 
for a desktop machine (although still feasible with modern supercomputers).
For the SNAP or LSST surveys with $N \approx 10^8$, 
this algorithm will require almost a 
terabyte of memory.  This would be a lot by today's standards, but when 
the data from these surveys eventually become available, 
it will probably be quite manageable.  Likewise, this algorithm would take
a good fraction of a year for the desktop machine described above,
but in five or ten years, it will must faster.
In any case, this algorithm is extrapolated to be about 20 times faster than
the previous algorithm.

If anyone is interested in the code for any of these algorithms or 
would like to use our code on a survey dataset, we would
be happy to provide it for you. 
If interested, please send an email to Mike Jarvis (mjarvis@hep.upenn.edu).

\section{Application to Simulated and Observed Data}
\label{applicationsection}

\subsection{Simulated Shear}
\label{simsection}

To verify that the above equations are correct, we first apply them
to a simulated shear field produced by 
ray tracing simulations \citep*{JSW}.  The simulated data 
uniformly cover an area $200\arcmin$ on a side, and have
no shape noise on the shear values.  The particular simulation
we used was for $\Lambda$CDM with $\Omega_m = 0.3$, $\Lambda = 0.7$,
$\Gamma = 0.21$ and $\sigma_8 = 0.9$, with sources at $z = 1$.

We measure $\mapcb$ both from the correlation functions as described
above and also by measuring $\map$ for many apertures using
\eqq{mapdefq} and then computing $\mapcb$ directly.  This second
technique is not possible on real data, due to masking effects
from bright stars or complicated survey geometry, as discussed
in \S\ref{introduction}, but for the simulated data, there
is no masking, so the direct computation is possible.

\begin{figure}
\includegraphics[width=120mm]{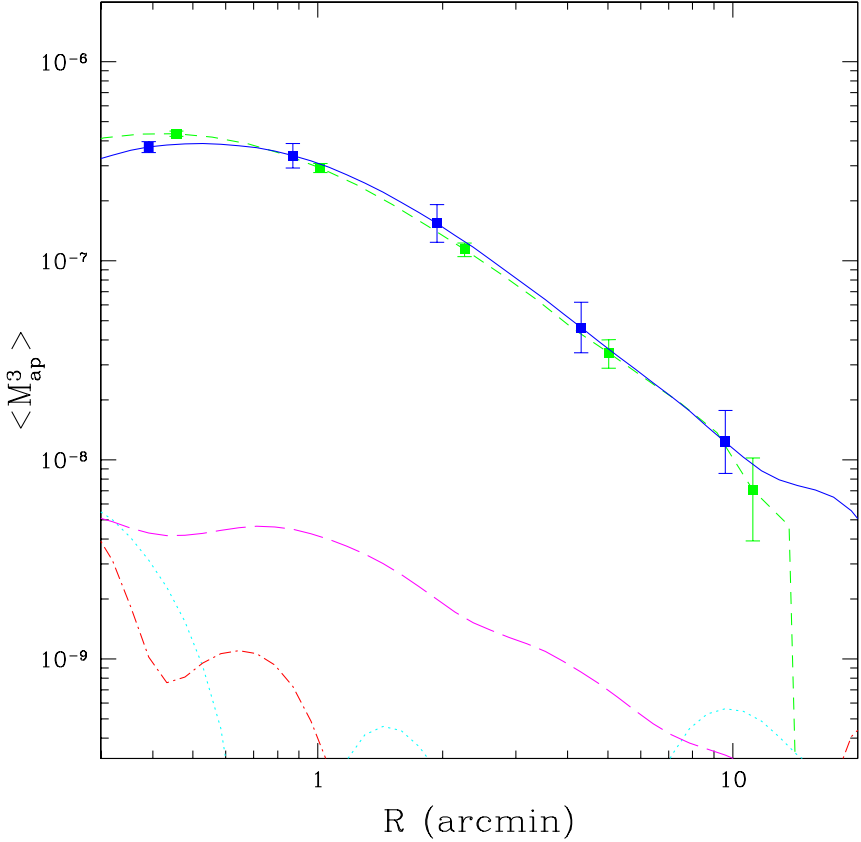}
\caption[]{\small
A comparison between a direct calculation of $\mapcb$ and the 
integration technique presented in this paper for a simulated 
shear field with no shape noise or masking.  The direct
calculation is the green dashed curve, and the integration
technique is the blue solid curve. 
The three other curves near (and often below)
the bottom of the plot are the absolute values of the mixed
and B-modes measured by the integration technique.  
}
\label{comparisonfigure}
\end{figure}

The results from the two methods are shown in 
Fig.~\ref{comparisonfigure}.  The results from the direct aperture 
method (\eqq{mapdefq}) are the dashed green curve.
The results from the integration method (\S\ref{3ptsection})
using $b = 0.1$ are the solid blue curve.  
The error bars for the direct 
method were estimated from the distribution of $M_{\rm ap}$ 
values used to compute the average.  For the integration 
technique, we took 12 samples of 100,000 points drawn from the 
gridded simulation field, and computed the error bars from the
actual `field-to-field' variance.  The sampling is needed 
to prevent the gridding signature of the simulation 
from showing up in the 
correlation functions and affecting the results.

The two methods are seen to agree to within
the error bars over most of the range, 
so we believe the formulae in \S\ref{3ptsection} to
be accurate.  The largest discrepency is at scales just above
$10\arcmin$ where the direct measurement abruptly falls off.
We think that this is due to there being too few apertures in 
the field to accurately calculate the skewness.  At 
$R = 10\arcmin$, $U_R(r)$ is significant out to $20\arcmin$, 
so one can only fit 5 non-overlapping apertures of this 
size within the $200\arcmin$ square.  We do use overlapping
apertures for our calculation, since the measurements should
still be nearly independent when the apertures overlap
somewhat,
but we believe that the relatively small number of independent apertures
is likely to be the source of the discrepency.

The figure also shows the mixed and B-modes measured
by the integration technique: $\mapsqmx$ is the cyan dotted curve,
$\mapmxsq$ is the magenta long dashed curve, and $\mxcb$ is the red
dot-dashed curve.  The plotted values are the 
absolute values of the measurements, and the largest of these
($\mapmxsq$) is roughly two orders of magnitude below the E-mode
signal.  This is due to the use of $b = 0.1$; using a
smaller value would reduce the leakage of power from the E-mode
to the others still further.

\subsection{CTIO Survey Data}
\label{ctiosection}

We next apply the above formulae to our 75 square degree CTIO
(Cerro Tololo Interamerican Observatory)
survey.  A complete description of the survey data and the 
processing techniques used is found in \citet{Ja03}.  
The 75 total square degrees are divided among 12 fields, each
roughly 2.5 degrees square.  Each field has of order 150,000 
usable galaxies. 
We have also corrected the error pointed out by
\citet{Hi03} in our dillution correction formula.  The 
data presented here use their linear approximation
for this correction, which they find to be significantly
more accurate than our formula.

\begin{figure}
\includegraphics[width=120mm]{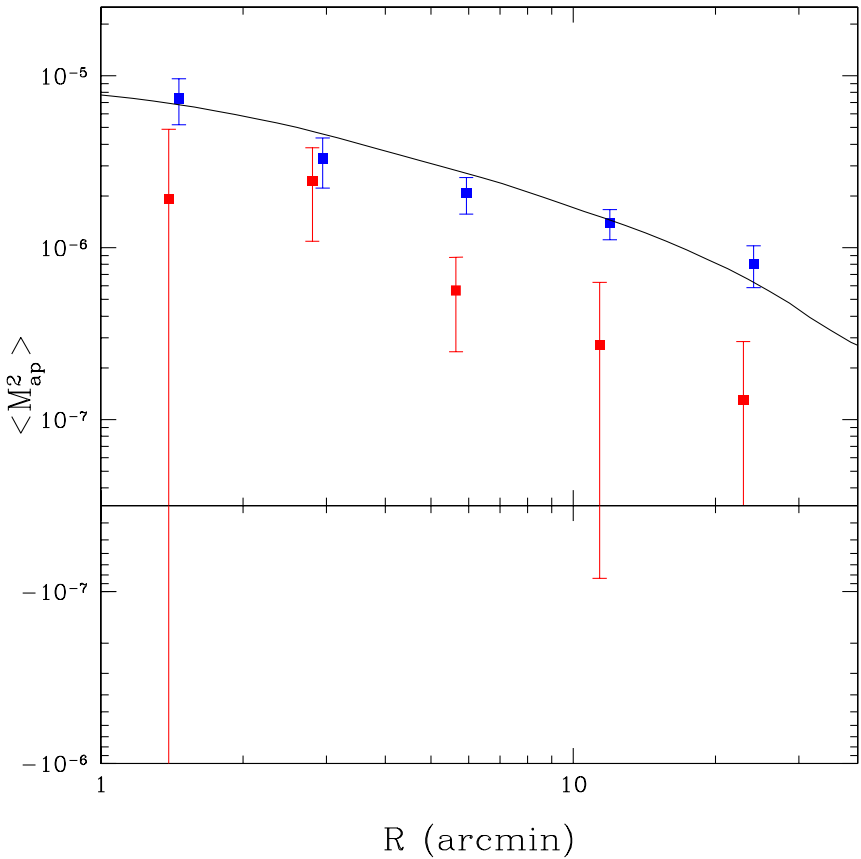}
\caption[]{\small
The variance of the aperture mass statistic using the 
integration technique for our CTIO 75 square degree survey.
The E-mode component is blue, and the B-mode component is red.
As discussed in \citet{Ja03}, there is evidently 
significant B-mode contamination at scales less than
$10\arcmin$.  The black curve is the theoretical
prediction for the concordance $\Lambda$CDM model.
}
\label{m2figure}
\end{figure}

In our previous paper, we measured the $\mapsq$ 
statistic and found a clear detection of the E-mode signal,
but found significant B-mode as well.  
We had used the aperture mass definition of \citet{Sch02} 
rather than the one defined herein (\eqq{mapdefu}), so we
present the results for this definition in 
Fig.~\ref{m2figure}.  The E-mode ($\mapsq$) is shown
in blue, and the B-mode ($\mxsq$) in red.
The measurements shown are spaced by approximately a 
factor of 2 in $R$, since this is where the measurements
become independent of each other.\citep{Sch02}. 
Note that the error bars are calculated
from field-to-field variation, so they accurately 
represent both statistical and sample variance.

\begin{figure}
\includegraphics[width=120mm]{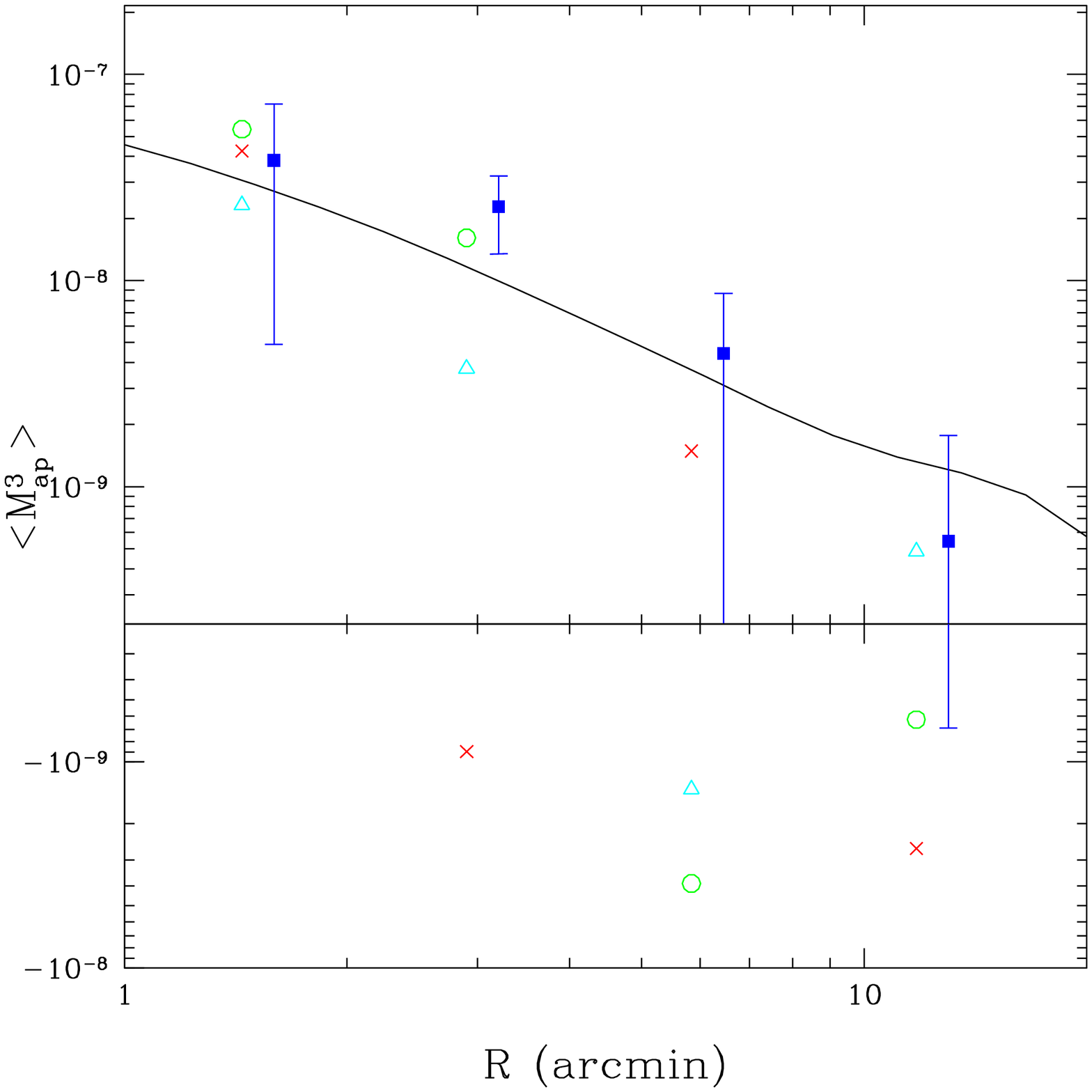}
\caption[]{\small
The skewness of the aperture mass statistic using the
integration technique for our CTIO 75 square degree survey.
The E-mode ($\mapcb$) is shown in blue, with 
error bars indicating the field-to-field
variance of the measurement.  The mixed and B-modes
are the circles ($\mapsqmx$), triangles ($\mapmxsq$),
and crosses ($\mxsq$).
The black curve is the theoretical prediction for the concordance
$\Lambda$CDM model.
}
\label{m3figure}
\end{figure}

Our calculation of the $\mapcb$ statistic, as well as the 
mixed and B-mode statistics, are shown 
in Fig.~\ref{m3figure} for the range of $1\arcmin
< R < 20\arcmin$.  Less than $1\arcmin$, the B-mode
contamination in the two-point measurement became large
compared to the signal.  And greater than $20\arcmin$,
we do not fully trust the measurement, since this is 
where the integration and direct measurements differed
for the simulated field described above.

With the integration technique, we can calculate
a value for $\mapcb$ at any value of $R$. 
However, as with the variance, we plot them spaced
by a factor of 2 in $R$, since this is when the 
points become uncorrelated.
Also, the error bars are again calculated 
from the field-to-field variation, so they inlude
sample variance.
We omit the error bars for the mixed and B-modes
for the sake of clarity, but their size is similar to 
those of the E-mode errors.  

Unfortunately, the detection is fairly marginal.  
The best conclusion we can draw is that $\mapcb$ is
roughly the right order of magnitude to be consistent 
with a concordance $\Lambda$CDM model, given
as the solid black curve in Fig.~\ref{m3figure}.  
(We scaled the measurements from the above simulation
to $\sigma_8 = 0.8$ and to $z_s = 0.6$ based on the 
theoretical scaling found by \citet*{Ta02}.)
However, the 
S/N is quite low, and the mixed and B-modes
indicate that there are potentially significant 
systematic errors contaminating the results.
Thus, we do not try to use this result to constrain 
any cosmological models.

One test we are willing to make with this data is to 
test whether it is consistent with zero.
The statistic we use for this test is:
\begin{equation}
P = \frac{1}{N} \sum {\frac{\mapcb(R_i)}{\sigma_i}}
\end{equation}
Over the range plotted, we find that 
$P = 1.09 \pm 0.48$\footnote{
The variance of P is $1/N$.  Since our integration
technique gives measurements essentially continuously 
within the range from $1\arcmin$ to $20\arcmin$, we calculate 
P using all of these points.  For the uncertainty,
we use the effective number of independent points:
$N = \ln(20)/\ln(2) = 4.3$.
}, which is a detection at the 
$2.3 \sigma$ level.  The other three modes give
values of 0.31, 0.14 and -0.13, all within $1 \sigma$
of zero.  So, we do have a mildly significant detection
that the skewness is positive, but we cannot make
any stronger claim than that.

\section{Conclusion}
\label{conclusion}

We have presented a new calculation for the skewness of
the aperture mass as an integral over the more easily
measured three-point correlation function.
We have also presented efficient algorithms for calculating
the three-point correlation function in real data.
These algorithms are believed to be faster than other 
published algorithms.  
Finally, we have applied these methods to our CTIO survey data,
for which we do not obtain a strong detection.  The signal
is consistent with the concordance model prediction, but
is only inconsistent with zero at the $2 \sigma$ level.

However, we expect that this method
will be useful for ongoing and future surveys which are 
deeper and wider than ours, and will therefore be better able 
to detect the signal.  
\citet{Pen03} have already used a similar technique on the 
VIRMOS/DESCART survey, and have measured an E-mode signal
which is significantly greater than the mixed and B-mode
contaminations (at least for some values of R).  They find
$\Omega_m < 0.5$ at 90 per cent confidence.

There are several ongoing and proposed surveys which will
substantially increase the S/N in these 
measurements.  For example, the Canada-France Legacy Survey
(CFLS) will cover
170 square degrees at a depth of $r < 25.7$.  
This is a similar depth to the VIRMOS/DESCART survey, but covers
about 20 times more area, so the S/N would be expected
to increase by a factor of 4.

The CFLS will observe three disjoint fields, two of 
49 square degrees and one of 72 square degrees,
with an estimated number of galaxies of roughly 600,000 and 900,000
respectively.  Thus, the total calculation time for our 
fastest three-point algorithm will be about 
3 days and require a maximum of about 7 GB of memory on 
a modern desktop computer, which is quite feasible.

In fact, this would calculate the correlation function for the 
entire range of scales available (up to 7 or 8 degrees), which
is not necessary.  Limiting the calculation to scales less than
about 200 arcmin, where the signal is strongest, speeds up the
calculation significantly.  In the current implementation,
this would not reduce the algorithm's memory demand, but 
one could easily modify the algorithm to keep only a fraction
of the total galaxies in memory at a time which would be useful
for surveys such as this one with large contiguous fields.

Improvements in the reduction techniques are probably just
as important as increasing the number of galaxies, since
the B-mode contamination is evidence that there are 
still some systematic errors in the data.  So far all 
surveys who have checked for B-modes in the data
show this contamination at a significant level, although
there are often scales at which the B-mode is small compared
to the E-mode signal.  Reducing these systematic effects will
probably require both better PSF-removal algorithms 
as well as cleaner raw images. 

\section*{Acknowledgments}

We thank Ue-li Pen, Masahiro Takada, Andrew Moore, Andrew Connolly,
Peter Schneider, and Martin Kilbinger
for useful discussions regarding the algorithms and results
presented here.
This work has been supported by grant AST-0320276 from the National
Science Foundation, NASA grant NAG5-10923, and a Keck foundation
grant.  We also thank the anonymous referee for a number of 
suggestions which have improved the paper.

\appendix
\section{Kurtosis}
\label{kurosissection}

The calculation of the fourth moment of the aperture mass can likewise 
be made from the four-point correlation function:
\begin{align}
\mqd(R) &= 
  24 \int \frac{s ds}{R^2} \int\limits_{s<t^\prime<|\bft^\prime-s|} \frac{d^2\bft^\prime}{2\pi R^2} 
  \int\limits_{t^\prime<u^\prime,t^\prime<|\bfu^\prime-s|} \frac{d^2\bfu^\prime}{2\pi R^2} 
  \Gamma^{(4)}_0(s,\bft^\prime,\bfu^\prime) 
  T^{(4)}_0\left(\frac{s}{R},\frac{\bft^\prime}{R},\frac{\bfu^\prime}{R}\right) \\
\mcbmc(R) &= 
  6 \int \frac{s ds}{R^2} \int\limits_{s<t^\prime<|\bft^\prime-s|} \frac{d^2\bft^\prime}{2\pi R^2} 
  \int\limits_{t^\prime<u^\prime,t^\prime<|\bfu^\prime-s|} \frac{d^2\bfu^\prime}{2\pi R^2} 
  \left[
    \Gamma^{(4)}_1(s,\bft^\prime,\bfu^\prime) 
    T^{(4)}_1\left(\frac{s}{R},\frac{\bft^\prime}{R},\frac{\bfu^\prime}{R}\right) \right. \nonumber\\
&\qquad \left.
    +\Gamma^{(4)}_2(s,\bft^\prime,\bfu^\prime) 
    T^{(4)}_2\left(\frac{s}{R},\frac{\bft^\prime}{R},\frac{\bfu^\prime}{R}\right) 
    +\Gamma^{(4)}_3(s,\bft^\prime,\bfu^\prime) 
    T^{(4)}_3\left(\frac{s}{R},\frac{\bft^\prime}{R},\frac{\bfu^\prime}{R}\right) %\right. \nonumber \\
%&\qquad \left.
    +\Gamma^{(4)}_4(s,\bft^\prime,\bfu^\prime) 
    T^{(4)}_4\left(\frac{s}{R},\frac{\bft^\prime}{R},\frac{\bfu^\prime}{R}\right) \right] \\
\msqmcsq(R) &= 
  8 \int \frac{s ds}{R^2} \int\limits_{s<t^\prime<|\bft^\prime-s|} \frac{d^2\bft^\prime}{2\pi R^2} 
  \int\limits_{t^\prime<u^\prime,t^\prime<|\bfu^\prime-s|} \frac{d^2\bfu^\prime}{2\pi R^2} 
  \left[
    \Gamma^{(4)}_5(s,\bft^\prime,\bfu^\prime) 
    T^{(4)}_5\left(\frac{s}{R},\frac{\bft^\prime}{R},\frac{\bfu^\prime}{R}\right) \right. \nonumber\\
&\qquad \left.
    +\Gamma^{(4)}_6(s,\bft^\prime,\bfu^\prime) 
    T^{(4)}_6\left(\frac{s}{R},\frac{\bft^\prime}{R},\frac{\bfu^\prime}{R}\right) 
    +\Gamma^{(4)}_7(s,\bft^\prime,\bfu^\prime) 
    T^{(4)}_7\left(\frac{s}{R},\frac{\bft^\prime}{R},\frac{\bfu^\prime}{R}\right) \right]
\end{align}

where we define the natural components of the four-point correlation function as did \citet{SL} 
(although our ordering is different):
\begin{align}
\Gamma^{(4)}_0(s,\bft^\prime,\bfu^\prime) &= \langle \gamma(\bfr) \gamma(\bfr+\bfs) \gamma(\bfr+\bft) \gamma(\bfr+\bfu) e^{-8i\alpha} \rangle \\
\Gamma^{(4)}_1(s,\bft^\prime,\bfu^\prime) &= \langle \gamma(\bfr)^* \gamma(\bfr+\bfs) \gamma(\bfr+\bft) \gamma(\bfr+\bfu) e^{-4i\alpha} \rangle \\
\Gamma^{(4)}_2(s,\bft^\prime,\bfu^\prime) &= \langle \gamma(\bfr) \gamma(\bfr+\bfs)^* \gamma(\bfr+\bft) \gamma(\bfr+\bfu) e^{-4i\alpha} \rangle \\
\Gamma^{(4)}_3(s,\bft^\prime,\bfu^\prime) &= \langle \gamma(\bfr) \gamma(\bfr+\bfs) \gamma(\bfr+\bft)^* \gamma(\bfr+\bfu) e^{-4i\alpha} \rangle \\
\Gamma^{(4)}_4(s,\bft^\prime,\bfu^\prime) &= \langle \gamma(\bfr) \gamma(\bfr+\bfs) \gamma(\bfr+\bft) \gamma(\bfr+\bfu)^* e^{-4i\alpha} \rangle \\
\Gamma^{(4)}_5(s,\bft^\prime,\bfu^\prime) &= \langle \gamma(\bfr)^* \gamma(\bfr+\bfs)^* \gamma(\bfr+\bft) \gamma(\bfr+\bfu) \rangle \\
\Gamma^{(4)}_6(s,\bft^\prime,\bfu^\prime) &= \langle \gamma(\bfr)^* \gamma(\bfr+\bfs) \gamma(\bfr+\bft)^* \gamma(\bfr+\bfu) \rangle \\
\Gamma^{(4)}_7(s,\bft^\prime,\bfu^\prime) &= \langle \gamma(\bfr)^* \gamma(\bfr+\bfs) \gamma(\bfr+\bft) \gamma(\bfr+\bfu)^* \rangle 
\end{align}

The corresponding $T$ functions are:
\begin{align}
T^{(4)}_0(s,\bft^\prime,\bfu^\prime) &= \frac{\bfq_1^{*2}\bfq_2^{*2}\bfq_3^{*2}\bfq_4^{*2}}{64}
\exp\left(-\frac{q_1^2 + q_2^2 + q_3^2 + q_4^2}{2}\right) \\
T^{(4)}_1(s,\bft^\prime,\bfu^\prime) &= \left[
\frac{\bfq_2^{*2}\bfq_3^{*2} + \bfq_2^{*2}\bfq_4^{*2} + \bfq_3^{*2}\bfq_4^{*2} -4\bfq_1^*\bfq_2^*\bfq_3^*\bfq_4^*}{128} 
+ \frac{\bfq_1 \bfq_2^* \bfq_3^* \bfq_4^* (\bfq_2^* \bfq_3^* + \bfq_2^* \bfq_4^* + \bfq_3^* \bfq_4^*)}{32} 
\right. \nonumber\\ &\qquad\left.
+ \frac{\bfq_1^2\bfq_2^{*2}\bfq_3^{*2}\bfq_4^{*2}}{64} \right]
\exp\left(-\frac{q_1^2 + q_2^2 + q_3^2 + q_4^2}{2}\right) \\
T^{(4)}_{2-4}(s,\bft^\prime,\bfu^\prime) &= T^{(4)}_1(s,\bft^\prime,\bfu^\prime) 
~\text{after a cyclic rotation of indices (1,2,3,4)} \\
T^{(4)}_5(s,\bft^\prime,\bfu^\prime) &= \left[
\frac{(2\bfq_1^*\bfq_2^* + (\bfq_1^* + \bfq_2^*)^2)(2\bfq_3^*\bfq_4^* + (\bfq_3^* + \bfq_4^*)^2)
-6 |\bfq_3 + \bfq_4|^2 + 3}{128}
\right. \nonumber\\ &\qquad\left.
- \frac{\bfq_1 \bfq_2 \bfq_3^* \bfq_4^* |\bfq_3 + \bfq_4|^2)}{32} 
+ \frac{\bfq_1^2\bfq_2^2\bfq_3^{*2}\bfq_4^{*2}}{64}\right]
\exp\left(-\frac{q_1^2 + q_2^2 + q_3^2 + q_4^2}{2}\right) \\
T^{(4)}_{6-7}(s,\bft^\prime,\bfu^\prime) &= T^{(4)}_5(s,\bft^\prime,\bfu^\prime) 
~\text{after a cyclic rotation of indices (2,3,4)}
\end{align}
where the $q$'s are again the vectors from each vertex to the centroid:
\begin{align}
\bfq_1 &= (s + \bft^\prime + \bfu^\prime)/4 \\
\bfq_2 &= (\bft^\prime + \bfu^\prime - 3s)/4 \\
\bfq_3 &= (s + \bfu^\prime - 3\bft^\prime)/4 \\
\bfq_4 &= (s + \bft^\prime - 3\bfu^\prime)/4
\end{align}

Finally, we can obtain the moments of \map\ and \mx\ from:
\begin{align}
\mapqd(R) &= \frac{1}{8} \Re \left( \mqd(R) + 4\mcbmc(R) + 3\msqmcsq(R) \right) \\ 
\mapcbmx(R) &= \frac{1}{8} \Im \left( \mqd(R) + 2\mcbmc(R) \right) \\ 
\mapsqmxsq(R) &= \frac{1}{8} \Re \left( -\mqd(R) + \msqmcsq(R) \right) \\ 
\mapmxcb(R) &= \frac{1}{8} \Im \left( -\mqd(R) + 2\mcbmc(R) \right) \\ 
\mxqd(R) &= \frac{1}{8} \Re \left( \mqd(R) - 4\mcbmc(R) + 3\msqmcsq(R) \right)
\end{align}

\end{document}